\documentclass[aps,prd,onecolumn,groupedaddress,preprintnumbers,superscriptaddress,showpacs]{revtex4}
\usepackage{amsmath,amssymb,latexsym}
\usepackage{graphicx}

\begin{document}

\title{
New branch of Kaluza-Klein compactification
}


\author{Shunichiro Kinoshita}
\email{kinoshita_at_utap.phys.s.u-tokyo.ac.jp}
\affiliation{   
Department of Physics, The University of Tokyo, Tokyo 113-0033, Japan   
} 

\preprint{UTAP-586} 
\preprint{RESCEU-86/07}

\date{\today}

\begin{abstract}
 We found a new branch of solutions in Freund--Rubin type flux
 compactifications.
 The geometry of these solutions is described as the external space
 which has a de Sitter symmetry and the internal space which is
 topologically spherical.
 However, it is not a simple form of $\mathrm{dS}_p\times S^q$ but a warped
 product of de Sitter space and a deformed sphere.
 We explicitly constructed numerical solutions for a specific case with
 $p=4$ and $q=4$.
 We show that the new branch of solutions emanates from the marginally
 stable solution in the branch of $\mathrm{dS}_4\times S^4$ solutions. 
\end{abstract}

\maketitle

\section{Introduction}

It is widely believed that de Sitter spacetime well describes our
accelerating Universe in the early and also in the present epoch.
An intriguing challenge is to realize the Universe as we know it 
in higher-dimensional theory such as string theory.
Therefore, it is interesting to obtain successful embeddings of
$4$-dimensional de Sitter space in higher-dimensional spacetimes.
In order to do so, it is necessary to stabilize the compact extra dimensions.
Without stabilization of the extra dimensions, it is impossible to obtain
effective $4$-dimensional theory in higher dimensional spacetimes.

Freund--Rubin compactification is simple model with a stabilization
mechanism by flux~\cite{Freund:1980xh}.
In this model, considering $(p+q)$-dimensional spacetime, there is a
$q$-form flux field for stabilizing the $q$-dimensional compact space.
Moreover, introducing a bulk cosmological constant allows an external
de Sitter space and an internal manifold with positive curvature.
We obtain a $(p+q)$-dimensional product spacetime of $p$-dimensional de
Sitter space $\mathrm{dS}_p$ and $q$-dimensional sphere $S^q$. 

If the Hubble parameter is higher than the critical value determined by the
flux that stabilizes the extra dimensions, in the
scalar sector, massless mode appears at the critical value and this mode
eventually becomes tachyonic~\cite{Bousso:2002fi,Martin:2004wp}
(also see, for example,
\cite{Frolov:2003yi,Contaldi:2004hr,Kinoshita:2007ci}).
This unstable mode is the homogeneous mode, in other words, the mode
corresponding to the change of the radius of the extra dimensions.
Physically, this instability can be interpreted based on thermodynamic
arguments as follows.
For a given total flux integrated over the compact space, which is the
conserved quantity of the system, there are some spacetimes with
different configurations.
We expect that if one configuration is unstable for a fixed total flux,
another more stable configurations will exist for the same parameter.
Indeed, we define the entropy of the system as one quarter of the total
area of de Sitter horizon, and we can show unstable solutions where
tachyonic modes appear belonging to an entropically lower branch in a family
of solutions.
Moreover, the critical value at which the upper branch and lower branch
merge is nothing but the one at which a massless mode
appears and the spacetime becomes marginally stable.
Thus the onset of the thermodynamic instability exactly coincides with
that of the dynamical instability~\cite{Kinoshita:2007ci}.

However, it is known that other tachyonic modes emerge as the number of
the extra dimensions is larger~\cite{Bousso:2002fi,Martin:2004wp}.
These additional tachyonic modes correspond to inhomogeneous
Kaluza--Klein excitations with quadrupole and higher multi-pole modes
($l\ge 2$), which indicate deformations of the compact internal space.
In order to shed light on its nature and nonperturbative properties,
the close connection between thermodynamic and dynamical stability could
be expected to play an important role. 
This new instability suggests the existence of a new branch
of solutions at the point where a marginal massless mode appears, 
as well as for black strings the nonuniform string branch emanates
from the uniform string one at the Gregory--Laflamme (GL) 
point~\cite{Gregory:1987nb,Gubser:2001ac}.
(See e.g.~\cite{Kol:2004ww,Harmark:2007md} and references therein.)

The aim of this paper is to examine a new and nontrivial branch of
solutions in Freund--Rubin type flux compactifications, which have a de
Sitter space and the {\it warped} compact space with sphere topology
rather than a round sphere.
In Sec.~\ref{sec:Freund_Rubin} we review general Freund--Rubin flux
compactifications and show trivial solutions with configuration
$\mathrm{dS}_p \times S^q$.
In Sec.~\ref{sec:new_branch} we explore nontrivial warped
solutions and show such a numerical solution for $p=4$ and $q=4$
explicitly.

\section{Freund--Rubin solution}
\label{sec:Freund_Rubin}

We briefly review general Freund--Rubin flux compactifications
with $\mathrm{dS}_p \times S^q$ and dynamical stability of those
configurations.
The ($p+q$)-dimensional action is 
\begin{equation}
 I = \frac{1}{16\pi}\int \mathrm d^{p+q}x \sqrt{-g}
  \left(R - 2\Lambda - \frac{1}{q!}F_{(q)}^2\right),
\end{equation}
where $F_{(q)}$ is the $q$-form field for stabilizing the $q$-sphere and
$\Lambda$ is a $(p+q)$-dimensional cosmological constant.
The equations of motion are 
\begin{equation}
 G_{MN} = \frac{1}{(q-1)!}
  F_{ML_1 \cdots L_{q-1}}F_N{}^{L_1 \cdots L_{q-1}}
  - \frac{1}{2q!}F_{(q)}^2 g_{MN} - \Lambda g_{MN},\label{eq:Einstein_eq}
\end{equation}
and
\begin{equation}
 \nabla_M F^{MN_1\cdots N_{q-1}} = 0\label{eq:Maxwell_eq}.
\end{equation}

For the trivial solution of Eqs.~(\ref{eq:Einstein_eq}) and
(\ref{eq:Maxwell_eq}), the metric and the $q$-form flux are given by 
\begin{equation}
 \mathrm ds^2 = - \mathrm dt^2 + e^{2ht}\mathrm d\vec{x}_{p-1}^2 +
  \rho^2 \mathrm d\Omega_q^2,\label{eq:metric_FR}
\end{equation}
and
\begin{equation}
 F_{(q)} = b \epsilon_{\mu_1 \cdots \mu_q},\label{eq:flux_FR}
\end{equation}
where $\epsilon_{\mu_1 \cdots \mu_q}$ is the volume element of the
$q$-sphere with a radius $\rho$.
Also $h$ and $b$ are the Hubble parameter of $p$-dimensional de Sitter
space and the flux strength, respectively.
From (\ref{eq:metric_FR}) and (\ref{eq:flux_FR}), the Einstein
and Maxwell equations reduce to two algebraic equations:
\begin{equation}
 \begin{aligned}
  (q-1)\rho^{-2} - (p-1)h^2 &= b^2,\\
  (q-1)^2\rho^{-2} + (p-1)^2h^2 &= 2\Lambda.
 \end{aligned}
 \label{eq:algebraic_relations}
\end{equation}
From the above equations, we obtain a relation in terms of two
parameters $b$ and $h$ when $\mathrm{dS}_p \times S^q$ solutions exist:  
\begin{equation}
 (p-1)(p+q-2)h^2 + (q-1)b^2 = 2\Lambda.\label{eq:b_h_relation}
\end{equation}

The analysis of linear perturbations in this background spacetime show
that there are two channels of instabilities in the scalar sector with
respect to the $p$-dimensional de Sitter symmetry:
(i) homogeneous excitation and (ii) inhomogeneous excitation with
quadrupole and higher multi-pole moments of Kaluza--Klein
modes~\cite{Bousso:2002fi,Martin:2004wp}.
(i) $l=0$ mode is a so-called volume modulus and becomes unstable as the
Hubble parameter $h$ of the de Sitter space becomes very large (i.e.,
the flux $b$ small).
This seems to be a generic feature of de Sitter
compactifications~\cite{Frolov:2003yi,Contaldi:2004hr}.
The stability condition that mass squared of the $l=0$ mode will not be
tachyonic is given by 
\begin{equation}
 h^2 \le \frac{2\Lambda(p-2)}{(p-1)^2(p+q-2)} \quad \text{or} \quad
  b^2 \ge \frac{2\Lambda}{(p-1)(q-1)}.\label{eq:condition_l0}
\end{equation}
Furthermore, as mentioned in the introduction, we can derive this
threshold according to the thermodynamic argument using
entropy~\cite{Kinoshita:2007ci}.

On the other hand, instabilities arising from (ii) $l \ge 2$ modes
appear when the number of extra dimensions is higher, to be precise,
$q\ge 4$.
The stability condition that the $l=2$ tachyonic mode should not appear
is  
\begin{equation}
 h^2 \ge 
  \frac{2\Lambda[2+q-3pq+(p-1)q^2]}
  {q(q-3)(p-1)^2(p+q-2)} \quad \text{or} \quad 
  b^2 \le \frac{4\Lambda}{q(q-3)(p-1)}.
  \label{eq:condition_l2}
\end{equation}

Finally, let us summarize the stability of Freund--Rubin solutions with
$\mathrm{dS}_p\times S^q$.
For $q=2$ and $q=3$ there is only one channel of instability arising from
$l=0$ mode.
The solutions for fluxes higher than the critical value given by 
(\ref{eq:condition_l0}) are stable.
For $q=4$, an additional tachyonic mode for $l=2$ appears and stable
configurations are allowed in the range of (\ref{eq:condition_l0}) and
(\ref{eq:condition_l2}).
For $q \ge 5$, there is no stable $\mathrm{dS}_p\times S^q$ solution 
for any value of fluxes.

\section{New branch}
\label{sec:new_branch}

In this section we explore a warped solution such that the $p$-dimensional
external space have a de Sitter symmetry with a warp factor depending
on internal coordinates and the $q$-dimensional internal space is
a deformed sphere.
The $(p+q)$-dimensional geometry of such solutions is generally
described as the metric form 
\begin{equation}
 \mathrm ds^2 = A(y)^2[ - \mathrm dt^2 + e^{2ht}
  \mathrm d\vec{x}_{p-1}^2]
 + g_{ab}(y)\mathrm dy^a\mathrm dy^b,  
\end{equation}
where $A(y)^2$ is the warp factor depending on internal coordinates
$y^a$, and $g_{ab}$ is the $q$-dimensional metric of the internal
compact manifold which is topologically $S^q$.
For simplicity, we assume that the internal space has $SO(q)$ isometry,
then the $(p+q)$-dimensional metric reduces to
\begin{equation}
 \mathrm ds^2 = A(r)^2[ - \mathrm dt^2 + e^{2ht}
  \mathrm d\vec{x}_{p-1}^2]
  + B(r)^2\mathrm dr^2 + C(r)^2\mathrm d\Omega_{q-1}^2.
\end{equation}
As there is still coordinate freedom in this metric, we can choose 
\begin{equation}
 A(r) = e^{\Phi(r)}, \quad B(r) = e^{-\frac{p}{q-2}\Phi(r)}, \quad 
  C(r) = e^{-\frac{p}{q-2}\Phi(r)}a(r).
\end{equation}
Note that this choice makes the problem similar to that of 
Friedmann--Robertson--Walker cosmology
with a scalar field (see appendix and for an example, \cite{Felder:2001da}).
Consequently, the metric form is given by 
\begin{equation}
 \mathrm ds^2 = e^{2\Phi(r)}[- \mathrm dt^2 + e^{2ht} 
  \mathrm d\vec x_{p-1}^2]
  + e^{-\frac{2p}{q-2}\Phi(r)}[\mathrm dr^2 + a^2(r) \mathrm d\Omega^2_{q-1}],
\end{equation}
and the $q$-form flux is
\begin{equation}
 F_{(q)} = b e^{-\frac{2p(q-1)}{q-2}\Phi}a^{q-1} \mathrm dr \wedge
  \mathrm d\Omega_{q-1},
\end{equation}
where $b$ is a constant.
This $q$-form field satisfies the Maxwell equation (\ref{eq:Maxwell_eq})
and the Bianchi identity $\mathrm dF_{(q)} = 0$.
It is worth noting that the values of two parameters $b$ and $h$ are not
always globally meaningful quantities, because physical quantities locally
observed depend on the warp factor.
Here we have taken the forms in order that meanings of $b$ and $h$ should
become the same as in the previous definitions in
Eqs.~(\ref{eq:metric_FR}) and (\ref{eq:flux_FR}) which is realized under
a trivial solution $\Phi=0$.


Now we consider $p=4$ and $q=4$ case for an explicit example.
This is because Freund--Rubin solutions for $q \ge 4$ have the instability
arising from higher multi-pole modes, as we have seen.
In this case the equations of motion are given by 
\begin{equation}
  \begin{aligned}
  \frac{a''}{a} &= - 6 \Phi'^2 + \frac{{a'}^2 - 1}{a^2},\\
  \Phi'' &= 3h^2 e^{-6\Phi} + \frac{1}{2} b^2 e^{-12\Phi}
   - \frac{1}{3}e^{- 4\Phi} - 3\frac{a'}{a}\Phi' \\
   &= -3\Phi'^2 + \frac{1}{4}b^2 e^{-12\Phi}
   + \frac{1}{6}e^{- 4\Phi} - 3\frac{a'}{a}\Phi'
   + \frac{3}{2}\frac{{a'}^2 - 1}{a^2},\label{eq:OEDs}
 \end{aligned}
\end{equation}
where the prime denotes the derivative with respect to $r$.
Note that we rescaled $b \to b \Lambda^{1/2}$, $h \to h \Lambda^{1/2}$, 
$a \to a \Lambda^{-1/2}$ and $r \to r \Lambda^{-1/2}$ to normalize
$\Lambda$ to unity.
The constraint equation is given by 
\begin{equation}
 3\left[\left(\frac{a'}{a}\right)^2 -
		      \frac{1}{a^2}\right] 
 = 6 \Phi'^2
 + 6h^2 e^{-6\Phi}
 + \frac{b^2}{2}e^{-12\Phi}
 - e^{-4\Phi}.\label{eq:constraint}
\end{equation}

In order to solve these equations we will require boundary
conditions.
We assume the internal space is topologically spherical and $a=0$
represents poles of the sphere.
Hence, at a pole ($a=0$) boundary conditions are imposed to ensure that it is
regular, and at equatorial plane other boundary conditions are imposed
for the internal space to be symmetric with respect to this plane.
We note that it is an assumption for the internal space to be symmetric. 
This is because analyses of linear perturbations imply that
an inhomogeneous unstable mode first appears at a quadrupole ($l=2$)
mode, so that we expect to find a corresponding solution to the $l=2$
mode in linear perturbations.

We suppose that the equatorial plane is located at $r=0$, then
boundary conditions at $r=0$ are given by $\Phi(0) = 0$, $\Phi'(0) = 0$ and
$a'(0) = 0$.
Note that although a value of $\Phi(0)$ is arbitrary, we have set
$\Phi(0)=0$ for simplicity here.
Moreover, from the constraint equation (\ref{eq:constraint}) we
obtain a value of $a$ at $r=0$:
\begin{equation}
 a(0) = \sqrt{\frac{6}{2 - b^2 -12h^2}}\label{eq:initial_a}.
\end{equation}

Changing values of $b$ and $h$ as shooting parameters, we numerically
integrate the ODEs of (\ref{eq:OEDs}) from $r=0$ to $r>0$ to obtain the
desired solutions satisfying the boundary conditions using the shooting
method.
As we have mentioned, since we are looking for solutions which have
regular compact internal space, at the other side we require boundary
conditions $a=0$, $a'=1$ and $\Phi'=0$ from regularity at the pole.

\begin{figure}[t]
 \begin{center}
  \includegraphics[width=.48\linewidth]{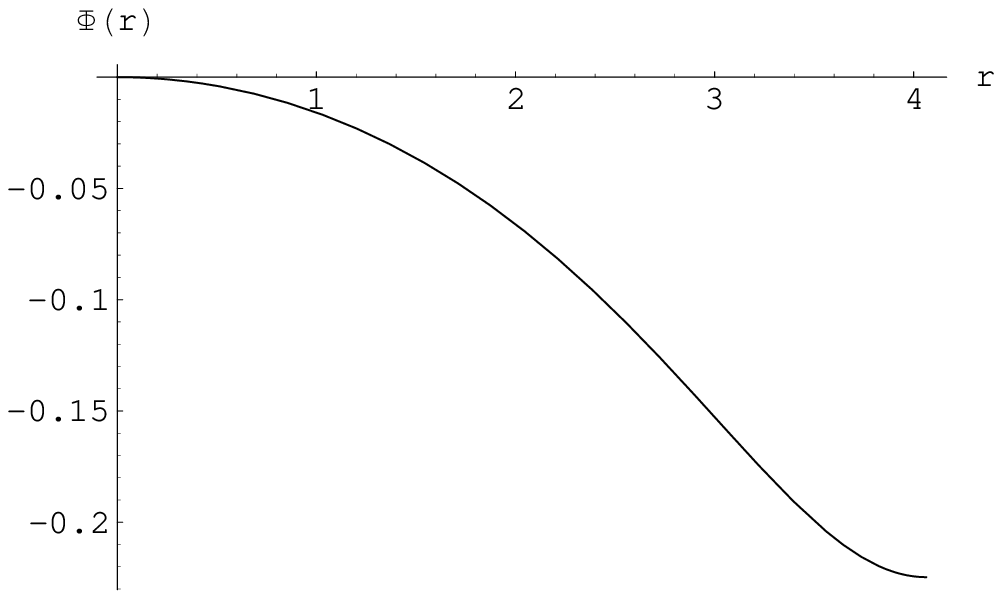}
  \includegraphics[width=.48\linewidth]{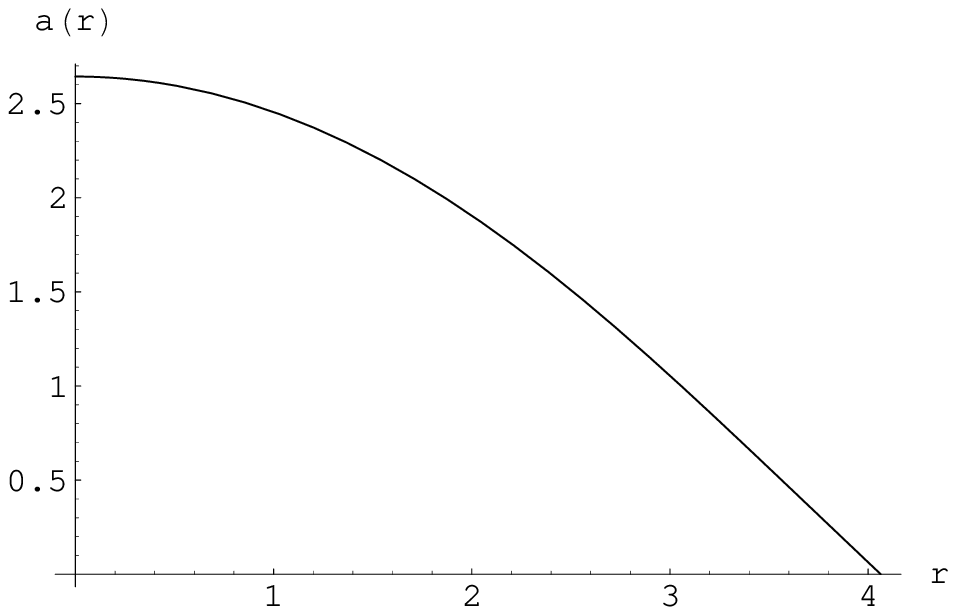}
  \caption{The numerical solutions of $\Phi$ (left) and $a$ (right) when
  $b=1/\sqrt{15}$ and $h=0.29928$. $r=0$ represents the equatorial plane
  and the pole is located at $r$ where $a(r)=0$.}
  \label{fig:numerical_solution1}
 \end{center}
\end{figure}

\begin{figure}[t]
 \begin{center}
  \includegraphics[width=.48\linewidth]{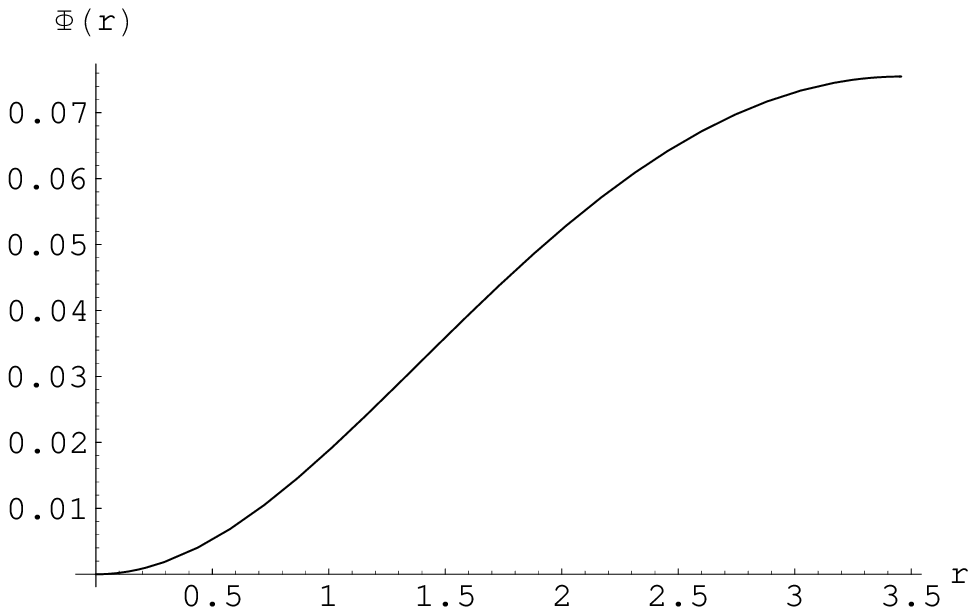}
  \includegraphics[width=.48\linewidth]{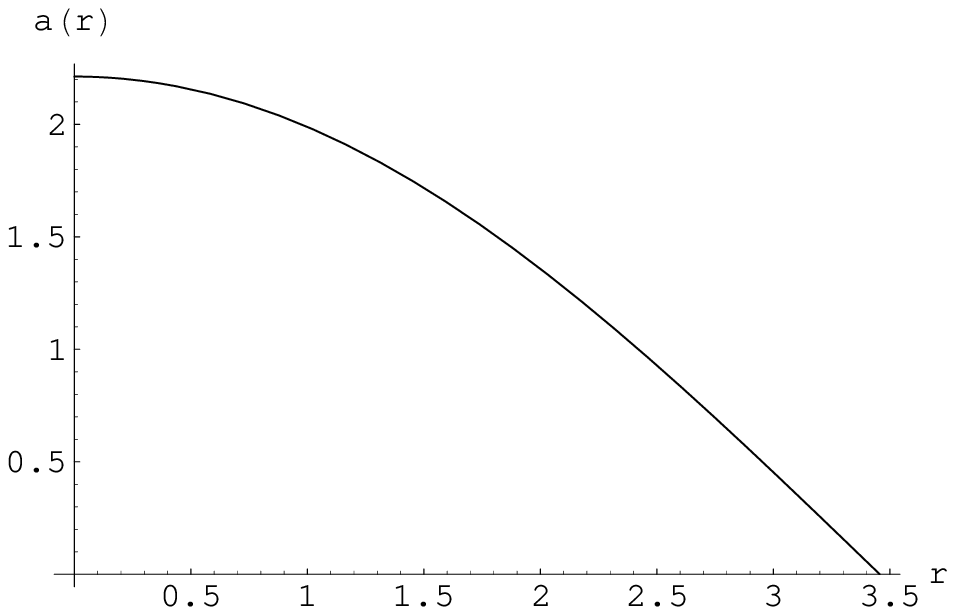}
  \caption{The numerical solutions of $\Phi$ (left) and $a$ (right) when
  $b=\sqrt{11/15}$ and $h=0.058325$.}
  \label{fig:numerical_solution2}
 \end{center}
\end{figure}

\begin{figure}[t]
 \begin{center}
  \includegraphics[width=.6\linewidth]{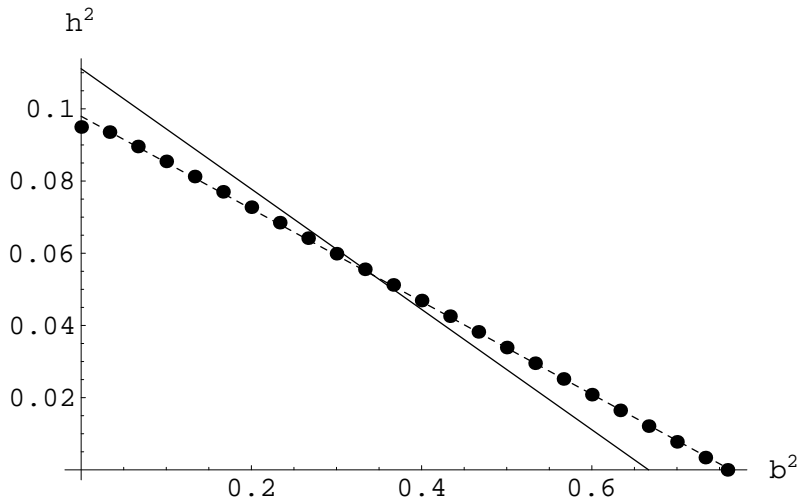}
  \caption{Two branches of solutions in the ($b^2,h^2$) space. The solid
  line represents the branch of Freund--Rubin solutions. The bold points
  correspond to values calculated numerically for which warped solutions
  have been found. It is likely that the branch of warped solutions
  would have a linear relation (dashed line) between $b^2$ and
  $h^2$ also. }
  \label{fig:b_h}
 \end{center}
\end{figure}

As a result, we have two families of solutions.
One is a family of well-known, trivial solutions given by 
\begin{equation}
\Phi(r) = 0, \quad a(r) = \rho\cos\frac{r}{\rho}
\end{equation}
where $\rho = a(0)$ from (\ref{eq:initial_a}) with satisfying 
the relation $18h^2 + 3b^2 = 2$ to which Eq.~(\ref{eq:b_h_relation}) leads.
The other is a family of nontrivial solutions.
For example, numerical solutions for some values of $b$ and $h$ are shown in
Fig.~\ref{fig:numerical_solution1} and~\ref{fig:numerical_solution2}.
Clearly $\Phi(r)$ is not constant, so that this solution is a warped one.
Figure~\ref{fig:b_h} shows two families of solutions in terms of two
parameters $b$ and $h$.
We see from this figure that the branch of trivial solutions and that of
nontrivial, warped solutions intersect at one point 
$(b^2,h^2) = \left(\frac{1}{3}, \frac{1}{18}\right)$, where there is
only a trivial solution.
What is especially important is that a massless mode for $l=2$ appears
on this point which is given by (\ref{eq:condition_l2}) according to
linear perturbations.
This means that the new branch of warped solutions emanates from the marginally
stable unwarped solution as we have expected.

\section{conclusion}
\label{sec:conclusion}

In this paper we have studied new solutions in Freund--Rubin type flux
compactifications.
By numerical analysis we found warped solutions with a deformed sphere
of the internal space. 
A new branch of these solutions intersects the branch of Freund--Rubin
solutions at a point in $(b^2,h^2)$-space.
The parameters represented by this point agree with those of the
spacetime configuration that massless mode for $l=2$ in the scalar
sector emerges under linear perturbations.
This is very similar to the case of black strings in which the uniform string
branch and the nonuniform branch meet at the GL point where the GL
instability occurs.
Therefore, it seems quite probable that for higher multi-pole mode
instability there is the close connection between thermodynamic and
dynamical stability.
We hope that there are rich phase (thermodynamic) structures not only
for black objects, but for de Sitter compactifications.
Of course, we do not know the stability of new branch of solutions yet.
However, the existence of another branch of solutions implies that
unstable configurations for large values of fluxes may be stabilized by
deformation of the internal geometry and change of flux distribution on
the sphere.
Also, even if the number of the extra dimensions is too large for stable
$\mathrm{dS}_p\times S^q$ solutions to exist, there may be stable
configurations in the warped branch.
We will reveal thermodynamic properties of this system in a forthcoming work.
It would provide evidence that a wider class of gravitating systems other
than black objects can be applied to the {\it correlated stability
conjecture}~\cite{Gubser:2000mm,Gubser:2000ec}.

\begin{acknowledgments}
 We gratefully acknowledge discussions with Shinji Mukohyama.
 The work was in part supported by JSPS through a Grant-in-Aid.
\end{acknowledgments}

\appendix

\section{}


The equations of motion are 
\begin{equation}
 \begin{aligned}
  \frac{a''}{a} &= \frac{b^2}{q - 2}
  e^{- \frac{2p(q - 1)}{q - 2}\Phi}
 + 
  \frac{p(p - 1)}{q - 2}
  h^2 e^{-\frac{2(p + q - 2)}{q - 2}\Phi}
  - 
  \frac{2\Lambda}{q - 2} e^{- \frac{2p}{q - 2}\Phi}
  -
   (q - 2)\frac{{a'}^2 - 1}{a^2}\\
  &= - \frac{p(p+q-2)}{(q-2)^2} \Phi'^2 + \frac{{a'}^2 - 1}{a^2},\\
  \Phi'' &= (p - 1)h^2
   e^{-\frac{2(p + q - 2)}{q - 2}\Phi}
   + \frac{q - 1}{p + q - 2}
  b^2 e^{- \frac{2p(q - 1)}{q - 2}\Phi}
  - 
  \frac{2\Lambda}{p + q - 2}e^{- \frac{2p}{q - 2}\Phi} - 
  (q - 1)\frac{a'}{a}\Phi' .
 \end{aligned}
\end{equation}
The constraint equation is given by 
\begin{equation}
 \frac{(q-1)(q-2)}{2}\left[\left(\frac{a'}{a}\right)^2 -
		      \frac{1}{a^2}\right] 
 = \frac{p(p+q-2)}{2(q-2)} \Phi'^2
 + \frac{p(p-1)}{2}h^2 e^{-\frac{2(p+q-2)}{q-2}\Phi}
 + \frac{b^2}{2}e^{-\frac{2p(q-1)}{q-2}\Phi}
 - \Lambda e^{-\frac{2p}{q-2}\Phi}.
\end{equation}

This system of equations can be rewritten as follows:
\begin{equation}
 \begin{aligned}
  \frac{\mathrm d\Phi}{\mathrm dr} &= \Psi,\\
  \frac{\mathrm d\Psi}{\mathrm dr} &= 
  - (q - 1)\mathcal H\Psi - \frac{q-2}{p(p + q - 2)}V_{,\Phi}(\Phi),\\
  \frac{\mathrm d\mathcal H}{\mathrm dr} &=
  - \mathcal H^2  
  - \frac{p(p+q-2)}{(q-1)(q-2)} \Psi^2 + \frac{2}{(q-1)(q-2)}V(\Phi),
 \end{aligned}
\end{equation}
where we define ``Hubble parameter'' as  
\begin{equation}
 \mathcal H \equiv \frac{a'}{a},
\end{equation}
and ``potential'' as 
\begin{equation}
 V(\Phi) \equiv \frac{p(p-1)}{2}h^2 e^{-\frac{2(p+q-2)}{q-2}\Phi}
 + \frac{b^2}{2}e^{-\frac{2p(q-1)}{q-2}\Phi}
 - \Lambda e^{-\frac{2p}{q-2}\Phi}.
\end{equation}
Also, the constraint equation becomes 
\begin{equation}
 \frac{(q-1)(q-2)}{2}\mathcal H^2  
  - \frac{p(p+q-2)}{2(q-2)} \Psi^2 - V(\Phi) =
  \frac{(q-1)(q-2)}{2}\frac{1}{a^2},
\end{equation}
which we can regard as ``Friedmann equation'' with a scalar field $\Phi$.

%
%
%



\begin{thebibliography}{99}

\bibitem{Freund:1980xh}
  P.~G.~O.~Freund and M.~A.~Rubin,
  Phys.\ Lett.\  B {\bf 97}, 233 (1980).

\bibitem{Bousso:2002fi}
  R.~Bousso, O.~DeWolfe and R.~C.~Myers,
  Found.\ Phys.\  {\bf 33}, 297 (2003)
  [arXiv:hep-th/0205080].

\bibitem{Martin:2004wp}
  J.~U.~Martin,
  JCAP {\bf 0504}, 010 (2005)
  [arXiv:hep-th/0412111].

\bibitem{Frolov:2003yi}
  A.~V.~Frolov and L.~Kofman,
  Phys.\ Rev.\ D {\bf 69}, 044021 (2004)
  [arXiv:hep-th/0309002].

\bibitem{Contaldi:2004hr}
  C.~R.~Contaldi, L.~Kofman and M.~Peloso,
  JCAP {\bf 0408}, 007 (2004)
  [arXiv:hep-th/0403270].

\bibitem{Kinoshita:2007ci}
  S.~Kinoshita, Y.~Sendouda and S.~Mukohyama,
  JCAP {\bf 0705}, 018 (2007)
  [arXiv:hep-th/0703271].

\bibitem{Gregory:1987nb}
  R.~Gregory and R.~Laflamme,
  Phys.\ Rev.\  D {\bf 37}, 305 (1988).

\bibitem{Gubser:2001ac}
  S.~S.~Gubser,
  Class.\ Quant.\ Grav.\  {\bf 19}, 4825 (2002)
  [arXiv:hep-th/0110193].

\bibitem{Kol:2004ww}
  B.~Kol,
  Phys.\ Rept.\  {\bf 422}, 119 (2006)
  [arXiv:hep-th/0411240].

\bibitem{Harmark:2007md}
  T.~Harmark, V.~Niarchos and N.~A.~Obers,
  Class.\ Quant.\ Grav.\  {\bf 24}, R1 (2007)
  [arXiv:hep-th/0701022].

\bibitem{Felder:2001da}
  G.~N.~Felder, A.~V.~Frolov and L.~Kofman,
  Class.\ Quant.\ Grav.\  {\bf 19}, 2983 (2002)
  [arXiv:hep-th/0112165].

\bibitem{Gubser:2000mm}
  S.~S.~Gubser and I.~Mitra,
  JHEP {\bf 0108}, 018 (2001)
  [arXiv:hep-th/0011127].

\bibitem{Gubser:2000ec}
  S.~S.~Gubser and I.~Mitra,
  arXiv:hep-th/0009126.

\end{thebibliography}
\end{document}